\RequirePackage{fix-cm}
\documentclass[smallextended]{svjour3}       
\smartqed  
\usepackage{graphicx}
\usepackage{mathptmx}      
\begin{document}

\title{
Effect of inhomogeneity of the Universe on 
a gravitationally bound local system: A no-go result for explaining 
the secular increase in the astronomical unit
}


\author{Hideyoshi Arakida
}


\institute{H. Arakida \at
Graduate School of Education, Iwate University
              \email{arakida@iwate-u.ac.jp}           
}

\date{Received: date / Accepted: date}

\maketitle

\begin{abstract}
We will investigate the influence of the inhomogeneity of the 
universe, especially that of the Lema{\^i}tre--Tolman--Bondi (LTB) 
model, on a gravitationally bound local system such as the solar system.
We concentrate on the dynamical perturbation to the planetary motion 
and derive the leading order effect generated from the LTB model. 
It will be shown that there appear not only a well-known
cosmological effect arisen from the homogeneous and isotropic model, 
such as the Robertson--Walker (RW) model, but also the additional 
terms due to the radial inhomogeneity of the LTB model. We will also 
apply the obtained results to the problem of secular increase in the 
astronomical unit, reported by Krasinsky and Brumberg (2004), 
and imply that the inhomogeneity of the universe 
cannot have a significant effect for explaining the observed 
$d{\rm AU}/dt = 15 \pm 4 ~{\rm [m/century]}$. 
\end{abstract}
\section{Introduction\label{intro}}
The advancements in astronomical and astrophysical measurement 
techniques, particularly those involving with the solar system, 
has been achieved remarkable accuracy of up to 9 to 11 digits level. 
These technical advancements have drastically improved the accuracy 
of planetary ephemerides such as DE \cite{standish2003}, EPM 
\cite{pitjeva2005}, VSOP \cite{bretagnon1988} and INPOP 
\cite{fienga2008} and that of various astronomical constants. 
With the increasingly improved measurement techniques, 
observational models are also required to be more accurate and 
rigorous; for details, refer to \cite{soffel2003} and the
references therein.

High-precision observational data also play a crucial role in 
experimental relativity \cite{will1993,will2006}. 
Presently, the main parameters of parametrized post-Newtonian (PPN) 
approximation, $\beta$ and $\gamma$ are tightly constrained to the 
value of general relativity, i.e., $\beta = \gamma = 1$. 
For a more accurate verification of gravity, space tests such as 
LISA \cite{lisa}, LATOR \cite{lator}, and ASTROD/ASTROD i \cite{astrod}
have been planned.

Thus far, theoretical developments in studies of the solar system 
have pertained to slow motion, slow rotation and weak field 
approximation, 
$g_{\mu\nu} = \eta_{\mu\nu} + h_{\mu\nu}, |h_{\mu\nu}| \ll 1$, 
where $\eta_{\mu\nu}$ is the static Minkowski metric and 
$h_{\mu\nu}$ is the perturbation. See \cite{dsx1,dsx2,dsx3,dsx4,bk1,bk2}. However, it is well known that 
our universe is expanding at an accelerated rate \cite{perlmutter1999}. 
Therefore, it is natural to consider the situation that the metric 
tensor, instead of the Minkowskian metric, asymptotically 
reaches for the expanding spacetime or the background Minkowskian 
metric $\eta_{\mu\nu}$ is replaced by the cosmological type. 
Several investigations have been conducted, which combine the 
local metric, e.g., the Schwarzschild spacetime or the barycentric 
celestial reference system adopted by IAU, with the global cosmological 
comoving coordinates
\cite{mcvittie,jarnefelt1,jarnefelt2,jarnefelt3,es1,es2,schucking,noerd,cooperstock,ks2004,faraoni,sereno,adkins2007,kopeikin2007,carrera}. 

The cosmological contribution to the local system has thus far been 
discussed based on the homogeneous and isotropic cosmological model,
i.e., the Robertson--Walker (RW) model.
However, inhomogeneous cosmological models have recently attracted 
considerable attention since these models can provide a possibility 
to explanation for the observed accelerated cosmic expansion without 
introducing the concept of dark energy. For instance, the 
luminosity-distance was investigated in
\cite{tomita2000a,tomita2000b,tomita2001a,tomita2001b,tomita2001c} 
based on the local void model and in \cite{iguchi2002} using the 
Lema{\^i}tre--Tolman--Bondi (LTB) model 
\cite{lemaitre1933,tolman1934,bondi1947,pk2006}. Moreover, 
Kasai (2007) \cite{kasai2007} re-analyzed the observed Type Ia 
supernovae data and proposed a phenomenological method to 
describe the large-scale inhomogeneity of the universe.

Therefore, it would be significant and interesting to investigate 
the influence of the inhomogeneity of the universe on 
the gravitationally bound local system. As far as we know, 
this issue has previously been examined by Gautreau (1984) 
\cite{gautreau} and Mashhoon {\it et al.} (2007) \cite{mashhoon}. 
Gautreau studied the special case of the LTB model, 
${\cal E}(r) = 0$, (see (\ref{LTB1})) with respect to his 
cosmological theory of the curvature coordinates\footnote{Gautreau 
does not start from the original form of the LTB model. 
However, Krasi{\'n}ski \cite{krasinski} suggested 
that the model by Gautreau corresponds to the sub case of the 
LTB model, ${\cal E}(r) = 0$.}. While,
Mashhoon {\it et al.} investigated the cosmological contribution
due to the LTB model as the tidal dynamics in the Fermi 
normal coordinate system.

With the remarkable improvements in the observations, it has been found 
that there exist the unexplained phenomena in theory within the solar 
system; the pioneer anomaly \cite{anderson}, the Earth fly-by anomaly 
\cite{anderson2}, the secular increase in the astronomical unit 
\cite{kb2004}, and the anomalous perihelion precession of Saturn 
\cite{iorio2009}. Presently, the origins of these anomalies are far 
from clear. Nonetheless, they may be attributable to some fundamental 
properties of gravitation 
(see \cite{lpd2008} and the references therein).

Among such phenomena, the secular increase in the astronomical unit 
is of concern to us. From the analysis of radiometric data, 
Krasinsky and Brumberg \cite{standish2005} discovered the positive 
secular trend in AU 
as\footnote{In this paper, cy refers to century as used in the 
study of Krasinsky and Brumberg (2004).}
\begin{equation}
\frac{d {\rm AU}}{dt} = 15 \pm 4~ {\rm [m/cy]},
\end{equation}
see also \cite{standish2005}. Recently, Pitjeva and Standish
evaluated $d{\rm AU}/dt \simeq 20 ~[{\rm m/cy}]$ \cite{pitjeva2009}.
These estimated values are approximately 100 times the error of
the present best-fit value of ${\rm AU}$ \cite{pitjeva2005}, 
\begin{equation}
\frac{1 ~{\rm [AU]}}{1 ~ [{\rm m}]} 
\equiv {\rm AU} = 1.495978706960 \times 10^{11} \pm 0.1.
\label{au1}
\end{equation}
This secular trend in AU was found by using following 
relation \cite{krasinsky2007}
\begin{equation} 
 t_{\rm theo} = \frac{d_{\rm theo}}{c}
  \left[
   \mbox{AU} + \frac{d\mbox{AU}}{dt}(t - t_0)
  \right]~~~{\rm [s]}, 
  \label{au2}
\end{equation}
where $t_{\rm theo}$ is the computed value of the round-trip 
time of the light/signal (theoretical value); $d_{\rm theo}$, the
interplanetary distance evaluated from the lunar-planetary 
ephemerides in the unit of ${\rm [AU]}$; 
$c$, the speed of light in vacuum; and $t_0$, the initial 
epoch of ephemerides. ${\rm AU}$ and $d{\rm AU}/dt$ are, 
respectively, the astronomical unit and its time variation.
$t_{\rm theo}$ is compared with the observed lapse time $t_{\rm obs}$.

The time dependent term in (\ref{au2}) cannot be correlated with 
any theoretical prediction; hence, several attempts have been made 
to explain this phenomenon, such as the effects of the 
cosmological expansion \cite{kb2004,mashhoon,arakida2009}, mass loss 
of the Sun \cite{kb2004,noerd2008}, the time variation of 
gravitational constant $G$ \cite{kb2004}, and the
influence of dark matter \cite{arakida2010}.
However, none of these attempts have thus far been successful.

Krasinsky and Brumberg (2004) pointed out that the 
inhomogeneity or non-uniformity of the universe 
may have a possible explanation for $d{\rm AU}/dt$; 
however, they did not provide evidence to support 
this hypothesis. Hence, it is important to verify 
this indication. Further a clarification of 
the observational difference between homogeneous and 
inhomogeneous cosmological models in the local dynamics is 
important in the field of modern cosmology.

In this paper, we will focus on the LTB solution as 
the inhomogeneous cosmological model and investigate its 
contribution to planetary motion. In section \ref{sec:fr}, 
we will summarize the dynamical perturbation based on the 
isotropic and homogeneous RW model and Robertson-McVittie 
(RM) model. Next, in section \ref{ltb}, we will derive the 
dynamical perturbation attributed to the LTB model. 
As an application of the obtained results, we will consider the 
secular increase in the astronomical unit, reported by 
Krasinsky and Brumberg (2004), in section \ref{dau}. 
Finally in section \ref{concl}, we will conclude the paper.
\section{Dynamical Perturbation in the RW and RM Models\label{sec:fr}}
Before discussing the dynamical perturbation in the LTB model, 
let us provide a brief overview of the planetary perturbation 
due to the Robertson--Walker (RW) model, and subsequently, due to the 
Robertson--McVittie (RM) model. Without loss of generality, 
we first consider the flat ($k = 0$) RW metric in the standard 
comoving form,
\begin{eqnarray}
 ds^2 = - c^2 dt^2 + 
  a^2(t)[dr^2 + r^2 (d\theta^2 + \sin^2\theta d\phi^2)],
  \label{FLRW-1}
\end{eqnarray}
where $a(t)$ is the scale factor. The equation of motion of 
a test particle can be expressed as \cite{weinberg,soffel1989},
\begin{eqnarray}
 \frac{d^2x^i}{dt^2}
  &=& 
   - \Gamma^i_{\mu\nu}\frac{dx^{\mu}}{dt}\frac{dx^{\nu}}{dt}
  + \frac{1}{c}\Gamma^{0}_{\mu\nu}
  \frac{dx^{\mu}}{dt}\frac{dx^{\nu}}{dt}\frac{dx^i}{dt}
  \nonumber\\
 &=&
  - c^2\Gamma^i_{00} - 2c\Gamma^i_{0j}v^j - \Gamma^i_{jk}v^j v^k
  + \frac{1}{c}
  \left(c^2 \Gamma^0_{00} + 2c\Gamma^0_{0j}v^j + 
  \Gamma^0_{jk}v^j v^k\right)v^i,
  \label{eom-1}
\end{eqnarray}
where $t$ is the coordinate time; $\Gamma^{\lambda}_{\mu\nu}$, 
the Christoffel symbol; and $v^i$ the coordinate velocity.
Restricting the equatorial motion to $\theta = \pi/2$, 
the equations of motion for $r$ and $\phi$ are given by
\begin{eqnarray}
 \frac{d^2 r}{dt^2} - r\left(\frac{d\phi}{dt}\right)^2
  &=& - 2 H \frac{dr}{dt},
    \label{eom-21}\\
 \frac{d}{dt}\left(r^2 \frac{d\phi}{dt}\right)
  &=& - 2 H r^2 \frac{d\phi}{dt},
  \label{eom-22}
\end{eqnarray}
where $H \equiv \dot{a}/a$ is the Hubble parameter.
If we introduce the proper or radial length $R$ as,
\begin{eqnarray}
 R \equiv r a(t),
  \label{rR}
\end{eqnarray}
then (\ref{eom-21}) and (\ref{eom-22}) are rewritten as,
\begin{eqnarray}
 \frac{d^2 R}{dt^2} - R\left(\frac{d\phi}{dt}\right)^2
  &=& \frac{\ddot{a}}{a}R,
  \label{eom-31}\\
 \frac{d}{dt}\left(R^2 \frac{d\phi}{dt}\right) &=& 0.
  \label{eom-32}
\end{eqnarray}
Hence, from the point of view of $R$, the motion of the test particle 
in RW spacetime is governed by
\begin{equation}
F^{\rm (RW)}_{R} = \frac{\ddot{a}}{a}R,\quad
 F^{\rm (RW)}_{\phi} = 0.
\end{equation}
Next, in order to observe the effect of cosmological expansion on 
the Newtonian gravity, we adopt the RM solution 
\cite{robertson,mcvittie},
\begin{eqnarray}
ds^2 = - 
\left[
\frac{1 - \frac{GM}{2 c^2 r a(t)}}{1 + \frac{GM}{2 c^2 r a(t)}}
\right]^2 
c^2 dt^2
+
\left[
1 + \frac{GM}{2 c^2 r a(t)}
\right]^4 a^2(t)
(dr^2 + r^2 d\Omega^2),
\label{rm1}
\end{eqnarray}
where $d\Omega^2 = d\theta^2 + \sin^2 \theta d\phi^2$,
$G$ is the Newtonian gravitational constant, and $M$ is the 
mass of the central gravitating body, i.e., the Sun. (\ref{rm1}) 
is expressed in the Newtonian or 1st order approximation as
\begin{eqnarray}
 ds^2 
 = -
  \left[
   1 - \frac{2GM}{c^2 r a(t)}
  \right]c^2 dt^2
  + a^2(t)
  (dr^2 + r^2 d\Omega^2).
  \label{rm2}
\end{eqnarray}
(\ref{rm2}) can be alternatively obtained from the cosmological 
perturbation theories \cite{irvine1965,kodama1984,tomita1991,mukhanov1992,shibata1995,dodelson2003,adkins2007},
\begin{eqnarray}
 ds^2 = - [1 + 2\Psi(t, \mbox{\boldmath $x$})]c^2dt^2
  + a^2(t)[1 + 2\Phi(t, \mbox{\boldmath $x$})]
  \delta_{ij}dx^i dx^j,
\end{eqnarray}
where $\Psi$ relates to the Newtonian gravitational potential,
$\Phi$ is the perturbation to the spatial curvature, and
$\delta_{ij}$ is the Kronecker's delta symbol.

From (\ref{eom-1}) and (\ref{rm2}), the equations of motion 
can be expressed as
\begin{eqnarray}
 \frac{d^2 r}{dt^2} - r\left(\frac{d\phi}{dt}\right)^2 
  &=& - \frac{GM}{r^2 a^3} - 2 H \frac{dr}{dt},
  \label{rm3-1}\\
 \frac{d}{dt}\left(r^2 \frac{d\phi}{dt}\right)
  &=& - 2 H r^2 \frac{d\phi}{dt}.
  \label{rm3-2}
\end{eqnarray}
Further, using (\ref{rR}), the coordinates of (\ref{rm3-1}) 
and (\ref{rm3-2}) are transformed into the proper 
coordinates as
\begin{eqnarray}
 \frac{d^2R}{dt^2} - R\left(\frac{d\phi}{dt}\right)^2
  &=& - \frac{GM}{R} + \frac{\ddot{a}}{a}R,
  \label{rm4-1}\\
  \frac{d}{dt}\left(R^2\frac{d\phi}{dt}\right) &=& 0.
  \label{rm4-2}
\end{eqnarray}
From (\ref{rm4-1}) and (\ref{rm4-2}), we find that in the Newtonian 
or 1st order approximation, the leading term of dynamical perturbation 
obtained from the RM model is the same as those generated from the 
RW model, $F^{\rm (RW)}_{R}$ and $F^{\rm (RW)}_{\phi}$.
\section{Dynamical Perturbation in the LTB Model\label{ltb}}
It is generally difficult to construct any cosmological model
containing a gravitating body because of the non-linearity of 
general relativity; however, the Robertson--McVittie (RM) model
is an exception. As shown in the previous section, it may be a
practical working hypothesis that the equation of motion due to 
both gravitating body and the cosmological effect can be 
phenomenologically determined by the linear combination of the 
Newtonian gravitational attraction and the cosmological effect 
evaluated in the cosmological background metric without 
considering the gravitating body. 

With the above assumption, let us obtain the cosmological 
perturbations attributed to the LTB model, which can be 
used to replace $F^{\rm (RW)}_{R}, F^{\rm (RW)}_{\phi}$ 
given in the previous section. 

The metric of LTB spacetime in the standard comoving form is 
given by \cite{lemaitre1933,tolman1934,bondi1947,pk2006},
\begin{eqnarray}
 ds^2 = -c^2 dt^2 + \frac{1}{1 + 2{\cal E}(r)}
  \left(\frac{\partial {\cal R}}{\partial r}\right)^2 dr^2
  + {\cal R}^2 d\Omega^2.
  \label{LTB1}
\end{eqnarray}
Here, ${\cal R}$ denotes the functions of $t$ and $r$, and
\begin{eqnarray}
 {\cal E}(r) &=& \frac{1}{2c^2}
  \left(\frac{\partial {\cal R}}{\partial t}\right)^2
  - \frac{{\cal M}(r)}{{\cal R}} - \frac{1}{6}\Lambda {\cal R}^2,
  \\
  {\cal M}(r) &=& \frac{4\pi G}{c^2}\int\rho (t, r){\cal R}^2 
  \frac{\partial {\cal R}}{\partial r}dr,
\end{eqnarray}
in which $\Lambda$ is the cosmological constant, $\rho(t, r)$ is 
the density of the cosmological pressureless particles\footnote{In 
the case of the LTB model, the energy-momentum tensor
is given by $T^{\mu\nu} = \rho(t, r) u^{\mu}u^{\nu}$, where
$\rho(t, r)$ is the density and $u^{\mu}$ is the 4-velocity.}, 
and ${\cal E}(r)$ and ${\cal M}(r)$ are the arbitrary functions of 
$r$. ${\cal E}(r)$ is the generalization of the curvature parameter 
$k$ in the RW model, and ${\cal M}(r)$ is the active gravitational mass 
that generates the gravitational field. It may be noted 
that ${\cal R}$ has the dimension of physical length, namely, 
the source area distance or the luminosity distance, 
while $r$ is a dimensionless coordinate value\cite{pk2006}.

Using (\ref{eom-1}) and (\ref{LTB1}), we obtain the 
equations of motion for $r$ and $\phi$ as 
\begin{eqnarray}
 \frac{d^2 r}{dt^2} &=&
  - \left[
     2 \frac{\partial^2 {\cal R}}{\partial t \partial r}\frac{dr}{dt}
     +
     \frac{\partial^2 {\cal R}}{\partial r^2}
     \left(\frac{dr}{dt}\right)^2
     - (1 + 2{\cal E})
     {\cal R}\left(\frac{d\phi}{dt}\right)^2
   \right]\frac{1}{\frac{\partial {\cal R}}{\partial r}}
  \nonumber\\
  & &+
  \frac{1}{1 + 2{\cal E}}\frac{d{\cal E}}{dr}
  \left(\frac{dr}{dt}\right)^2,
  \label{LTB2}\\
  \frac{d^2\phi}{dt^2} &=&
  -\frac{2}{{\cal R}} 
  \left[
   \frac{\partial {\cal R}}{dt} 
   +
   \frac{\partial {\cal R}}{\partial r}
   \frac{dr}{dt}
  \right]\frac{d\phi}{dt},
  \label{LTB3}
\end{eqnarray}
where we ignored the ${\cal O}(c^{-2})$ and higher order terms. 
When we put the flat RW limit 
${\cal R} \rightarrow R = ra(t), {\cal E} \rightarrow k = 0$,
(\ref{LTB2}) and (\ref{LTB3}) reduce to (\ref{eom-31}) and 
(\ref{eom-32}), respectively.

In order to relate $r$ to ${\cal R}$ explicitly, 
we suppose that the background LTB spacetime is regular at 
the origin $r = 0$ where the central body is located, and that 
the test particle, such as a planet, moves around $r = 0$; 
hence, the cosmological redshift $z$ in this area is sufficiently small, 
$z \ll 1$. Thus according to \cite{mashhoon}, we adopt the 
following expansion forms for ${\cal R}, {\cal E}$ and ${\cal M}$ 
around $r = 0$ as
\begin{eqnarray}
 {\cal R}(t,r) &=& r a(t)
  \left[
   1 + 
   \frac{1}{2}\frac{1}{a(t)}\Delta(t) r
   + {\cal O}(r^2)
  \right],\quad
  \Delta (t) = 
  \left.\frac{\partial^2 {\cal R}}{\partial r^2}\right|_{r=0}
  \ll 1,\label{eq:R}\\
 {\cal E}(r) &=&
  \frac{1}{2}\epsilon r^2 + 
  {\cal O}(r^3),\quad
  \epsilon = \left.\frac{d^2 {\cal E}}{d r^2}\right|_{r = 0}
  \ll 1,\label{eq:e}\\
 {\cal M}(r) &=&
  \frac{1}{6} m r^3 + {\cal O}(r^4),\quad
  m = \left.\frac{d^3 {\cal M}}{d r^3}\right|_{r=0} \ll 1,
  \label{eq:m}
\end{eqnarray}
in which the scale factor $a(t)$ is defined as,
\begin{eqnarray}
 a(t) \equiv 
  \left.
   \frac{\partial {\cal R}}{\partial r}
  \right|_{r=0}.
\end{eqnarray}
Using these equations, (\ref{LTB2}) and (\ref{LTB3}) are 
rewritten as
\begin{eqnarray}
 \frac{d^2{\cal R}}{dt^2} - {\cal R}
  \left(\frac{d\phi}{dt}\right)^2 &=& {\cal F}^{\rm (LTB)}_{\cal R},
  \label{LTB4-1}\\
 \frac{d}{dt}
  \left(
   {\cal R}^2 \frac{d\phi}{dt}
  \right) &=& {\cal F}^{\rm (LTB)}_{\phi},
  \label{LTB4-2}
\end{eqnarray}
where the leading-order dynamical perturbations, 
${\cal F}^{\rm (LTB)}_{\cal R}$ and 
${\cal F}^{\rm (LTB)}_{\phi}$ are expressed as
\begin{eqnarray}
 {\cal F}^{\rm (LTB)}_{\cal R} &=& 
  \left[
  \frac{\ddot{a}}{a} + 
  \left(\frac{1}{\Delta}\frac{d\Delta}{dt}\right)^2
  \right]{\cal R}
  - \frac{2\epsilon}{\Delta}
  \left[
   \frac{{\cal R} \dot{a}^2}{a^2}
   -
   \frac{\dot{a}}{a}\dot{\cal R}
  \right]\nonumber\\
 &=&
  \left[
  -q \left(\frac{\dot{a}}{a}\right)^2 + 
  \left(\frac{1}{\Delta}\frac{d\Delta}{dt}\right)^2
  \right]{\cal R}
  - \frac{2\epsilon}{\Delta}
  \left[
   \frac{{\cal R} \dot{a}^2}{a^2}
   -
   \frac{\dot{a}}{a}\dot{\cal R}
  \right], \label{LTB6} \\
 {\cal F}^{\rm (LTB)}_{\phi} &=& 0.
\end{eqnarray}
In (\ref{LTB6}), we used the standard relation in the RW model,
\begin{eqnarray}
 \frac{\ddot{a}}{a} = -q \left(\frac{\dot{a}}{a}\right)^2,
\end{eqnarray}
where $q$ is the deceleration parameter. The first term in 
(\ref{LTB6}) is $F^{\rm (RW)}_{R}$, and the second to fourth terms 
are corrections obtained from the LTB model. It may be considered 
that the second term in (\ref{LTB6}) is analogous to 
$F^{\rm (RW)}_{R} = - q (\dot{a}/a)^2$.

In (\ref{LTB6}), we must evaluate $\epsilon$ and $\Delta (t)$.
Since the observational cosmology indicates that our universe 
has a flat geometry, we can set $\epsilon = 0$.
$\Delta (t)$ may be in principle obtained from the 
modified luminosity-redshift relation \cite{pm1984},
\begin{eqnarray}
 d_{\rm L} = 
  c\left[\frac{z}{H} + \frac{z^2}{2H}(1 - q - C)\right],\quad
  C = \frac{1}{aH^2}\frac{d\Delta}{dt}.
  \label{LTB7}
\end{eqnarray}
Finally, following the assumption discussed in the 
beginning of this section, the equation of motion, attributed 
to both the gravitating body and the cosmological effect 
due to the LTB model, can be phenomenologically given by,
\begin{eqnarray}
 \frac{d^2{\cal R}}{dt^2} - {\cal R}
  \left(\frac{d\phi}{dt}\right)^2 &=& 
  - \frac{GM}{\cal R} + {\cal F}^{\rm (LTB)}_{\cal R},
  \label{LTB5-1}\\
 \frac{d}{dt}
  \left(
   {\cal R}^2 \frac{d\phi}{dt}
  \right) &=& {\cal F}^{\rm (LTB)}_{\phi}.
  \label{LTB5-2}
\end{eqnarray}
\section{Secular Increase in Astronomical Unit\label{dau}}
In this section, as an application of (\ref{LTB5-1}) and 
(\ref{LTB5-2}), let us consider the secular increase in the 
astronomical unit \cite{kb2004,standish2005,arakida2009}.
Krasinsky and Brumberg found from their analysis of planetary 
radar and martian orbiters/landers range data that 
the astronomical unit (AU) increases with respect to meters as 
$d{\rm AU}/dt = 15 \pm 4~[{\rm m/cy}]$. This secular trend 
cannot be related to any theoretical model, and thus far, 
the origin of this secular increase is far from clear. 

Krasinsky and Brumberg suggested one possibility, that the 
inhomogeneity of the universe may be an explanation for 
$d{\rm AU}/dt$. We consider this possibility in terms of the
LTB model. Because the current cosmological observations assert 
that the geometry of our Universe is flat, we choose $\epsilon = 0$.
In this case, the cosmological perturbation 
${\cal F}^{\rm (LTB)}_{\cal R}$ becomes
\begin{equation} 
{\cal F}^{\rm (LTB)}_{\cal R} =
\left[
- q \left(\frac{\dot{a}}{a}\right)^2
+
\left(\frac{1}{\Delta}\frac{d\Delta}{dt}\right)^2
\right]{\cal R}.
\end{equation}
In our approximation, $-q (\dot{a}/a)^2 {\cal R}$ or, 
equivalently, $(\ddot{a}/a){\cal R}$ is a dominant cosmological
effect; however, its contribution is considerably negligible
\cite{jarnefelt1,jarnefelt2,jarnefelt3,noerd,cooperstock,carrera,ks2004,sereno,faraoni,adkins2007,arakida2009}.
From the assumption (\ref{eq:R}), $\Delta$ can be considered 
as a small correction to the scale factor $a(t)$; its time 
variation $d\Delta/dt$ is also smaller. Further, it is known that 
the deviation of temperature in the observed cosmic microwave 
background (CMB) radiation is of the order of $10^{-5}$; 
hence, we may use the following,
\begin{equation}
\frac{- q \left(\frac{\dot{a}}{a}\right)^2}
{\left(\frac{1}{\Delta}\frac{d\Delta}{dt}\right)^2}
\approx 10^{-5}.
\end{equation}
It is, therefore, currently difficult to detect the 
cosmological contributions attributed not only to the RW model 
but also to the inhomogeneity of the universe. The inhomogeneity 
in the background cosmological matter distribution 
does not have a detectable effect and hence cannot explain the 
observed $d{\rm AU}/dt$. 
\section{Conclusions\label{concl}}
We investigated the cosmological influence according to the LTB 
model on a gravitationally bound local system such as the solar system.
We focused on planetary motion and obtained the leading-order 
dynamical perturbation generated from the LTB spacetime.
The obtained dynamical perturbations, especially (\ref{LTB5-1}),
contains the contribution attributed to the RW model and also 
the correction terms attributed to the LTB model. 
Moreover, we applied the obtained results to the secular increase 
in the astronomical unit \cite{kb2004} and confirmed that 
the effect of the inhomogeneity of the universe 
does not provide an explanation for $d{\rm AU}/dt$.

In spite of several attempts (see section \ref{intro}), the origin 
of $d{\rm AU}/dt$ is far from clear. It is now pointed out that the 
most plausible reason of $d{\rm AU}/dt$ is due to either the lack 
of calibrations in the internal delays of radio signals within 
spacecrafts or the complication of the modeling of solar corona. 
However, none of the explanations have thus far been successful; 
this issue should hence be explored using all possibilities. 
A re-analysis of $d{\rm AU}/dt$
incorporating new data sets is also expected.

Since the astronomical unit, as expressed in (\ref{au1}), is 
currently determined from the arrival time measurement of radar signals, 
we must investigate this problem in terms of light/signal propagation.
To this end, we need to construct a cosmological model 
that combines the LTB model with a gravitating body.
Of course, it may be difficult to detect the cosmological effect in
the solar system. Nonetheless, since the theoretical discussions 
pertaining to this issue are still unresolved, it is, from 
theoretical point of view, interesting to develop a 
rigorous physical model that matches the gravitational bound local 
system and the cosmological models and clarify
several assertions.

\begin{acknowledgements}
We would like to thank the referees for comments and suggestions.
We also acknowledge Prof. G. A. Krasinsky for providing 
information and comments regarding the AU issue. 
This work was partially supported by the Ministry of Education, 
Science, Sports and Culture, Grant-in-Aid, No. 21740193.
\end{acknowledgements}


\end{document}